\documentstyle[12pt,epsfig]{article}
\begin{document}
\title{\bf Time emergence by self--measurement in a quantum anisotropic universe.}
\author{ A. Camacho  $^{\circ}$
\thanks{email: acamacho@aip.de}~~and A. Camacho--Galv\'an $^{\diamond}$\thanks{email: abel@servidor.unam.mx}\\
$^{\diamond}$DEP--Facultad de Ingenier{\'\i}a,\\
Universidad Nacional Aut\'onoma de M\'exico, \\ 
$^{\circ}$Astrophysikalisches Institut Potsdam, \\
An der Sternwarte 16, D--14482 Potsdam, Germany.}
\bigskip
\bigskip
\date{}
\maketitle

\begin{abstract}
We begin this work calculating Halliwell's propagator in the case of a Mixmaster universe with small anisotropy. Afterwards in the context of the Decoherence Model we introduce in our system terms that comprise the self--measurement of the universe of this model by higher multipoles of matter. 
Analyzing self--measurement with the Restricted Path Integral Formalism we obtain Halliwell's modified propagator and find that a gauge invariant phy\-si\-cal time emerges as consequence of this process. The conditions leading to Wheeler--DeWitt dynamics are also obtained. 
The comparison of our results with those of the isotropic case will enable us to conclude that the number of conditions to be satisfied in order to have Halliwell's regime is in the anisotropic situation bigger than in an isotropic universe. 
We obtain also in terms of the parameters of the measurement process an expression for the threshold in time beyond which the scale factors of this model are meaningless. 
\end{abstract}
\bigskip
\bigskip
\bigskip

PACS number: 98.80 Hw
\newpage
\section{Introduction.}
\bigskip

The precise formulation of a quantum theory of gravity is yet not known. But there exist several approaches which try to construct such a theory. One of them is the canonical formalism and the basic idea in it is the intention to derive equations for wave functionals on an adequate configuration space. 
This is done by foliating the classical spacetime into spatial hypersurfaces and choosing the spatial metric as a canonical variable [1]. 
In this approach spacetime is no longer a fundamental concept, and its role is taken over by the space of all three--dimensional geometries, which is called superspace and acts as configuration space for the theory. 
The central kinematical quantity $\psi$ is a wave function defined on superspace and on matter degrees of freedom.

The classical diffeomorphism invariance of general relativity leads to the presence of constraints: the total Hamiltonian must vanish, then the wave functional obeys the Wheeler--De Witt equation, $H\psi = 0$. 
Since due to the uncertainty relations no spacetimes exist anymore at the level of quantum gravity, there is no time parameter available, with other words, the Wheeler--DeWitt equation is timeless, one may say that there is no time in quantum cosmology. 

There are several attempts to overcome this difficulty, among them we may find a probabilistic definition of time [2]. 

Another interesting proposal in this quest is based on the Decoherence Model (DM) [3]. 
This model analyzes the appearance of classical properties of a given system by proposing a new term in its Hamiltonian that considers the interaction between its collective and microscopical degrees of freedom, the latter ones are also called environment. 

This new term in the Hamiltonian destroys quantum interferences at macroscopical level and the proposed model claims that the description of any quantum system is incomplete if it does not comprise this interaction. With other words, the a\-ppea\-rance of classical properties in quantum theory is possible only if this interaction is considered.

The idea of time emergence as a consequence of the self--measurement process of the quantum universe is already an old one [4].

The essential point in the application of DM on the case of quantum cosmology is that gravity couples to all forms of energy, gravity is measured by matter and therefore a general superposition of gravitational quantum states is decohered [5].  
With other words, if we consider the continuous measurement of the quantum universe, then its dynamics may be modified in such a way that time arises. 
The role of measuring device is played by higher multipoles of matter [5], which describe density fluctuations and gravitational waves present in the universe. 
These higher multipoles may thereby be considered as the environment associated to the superspace variables of the model, which in this proposal play the role of collective variables.

We begin this work calculating Halliwell's propagator in the case of a Mixmaster universe with small anisotropy. Afterwards continuous measurements in the dynamics of our universe are considered and employing the Restricted Path Integral Formalism (RPIF) [6] Halliwell's modified propagator will be found. 
One of the advantages of RPIF is that it allows us to take into account the influence of the measuring device without knowing the actual scheme of measurement [7].

We will show that time emerges as a quantitative feature of our model, namely a gauge invariant physical time emerges as 
consequence of the self--measurement process of the universe. The conditions leading to Wheeler--DeWitt equation are also 
obtained. 
Comparing with the already known case of an isotropic universe [8], we will see that even a very small (but non--vanishing) 
anisotropy imposes a set of conditions on the validity region of Halliwell's propagator that is not present in the isotropic case.

This set renders a functional dependence between the features of the self--mea\-su\-re\-ment process and the size of the neighborhoods in the 3--Geometry in which Wheeler--DeWitt equation is valid.
This result is no surprise at all. Indeed, according to DM we may consider the elements of the spatial metric as collective variables [3], and the presence of anisotropy implies that, 
with respect to the isotropic situation, we have now more collective variables. Therefore, in the anisotropic case the term in the Hamiltonian that considers the interaction between environment and collective degrees of freedom could play a more decisive role in the dynamics of the system than in the isotropic situation. 

We obtain also an expression, in terms of the parameters of this self--measurement process, that provides a threshold in time beyond which the concept of scale factor lacks physical meaning.
\bigskip

\section{Propagation Amplitudes.}
\bigskip
\bigskip 
Let us consider the Mixmaster metric

\begin{equation}
ds^2 = -N^2d^2\tau + e^{2\alpha}(e^{2\beta})_{ij}\sigma^i\sigma^j,
\end{equation}

\noindent where $\beta_{ij}$ are the elements of a traceless diagonal matrix, $N, \alpha$ and $\beta_{ij}$ are functions only of $\tau$, with $\sigma^1 = cos\varphi d\theta + sin\varphi sin\theta d\phi$, $\sigma^2 = sin\varphi d\theta - cos\varphi sin\theta d\phi$, $\sigma^3 = d\varphi + cos\theta d\phi$ and $det(e^{2\beta}) = 1$. We have the geometry of a homogeneous but not isotropic sphere [9].

Here $\tau$ is an arbitrary parameter related to the foliation of the classical spacetime into spatial hypersurfaces, and if it suffers the action of a transformation, namely if we have $d\tau \rightarrow d\zeta = f(\tau)d\tau$, then invariance demands also the transformation of the lapse function, we must also carry out the transformation $N(\tau) \rightarrow M(\zeta) = {N(\tau)\over f(\tau)}$. 

The (3 + 1) decomposition of the metric is [1, 9] $g_{ij} = e^{2\alpha}e^{2\beta_{ij}}\delta_{ij}$, $N_i = 0$, $N^{\bot} = N^{-1}$ and $\pi^{ij} = {e^{\alpha - 2\beta_{ij}}\over N}(\dot{\beta}_{ij} -2\dot{\alpha)}\delta_{ij}$.

The action is 

\begin{equation}
S = \int (\pi^{ij}\dot{g}_{ij} - N^{\mu}H_{\mu})d^4x,
\end{equation}

\noindent we use $\hbar = 1$, $c = 1$ and $G = 1$.

For this particular case we have 

{\setlength\arraycolsep{2pt}\begin{eqnarray}
\pi ^{ij}\dot{g}_{ij} - N^{\mu}H_{\mu} =  {2\over N}e ^{3\alpha}\{(\dot{\beta}_{11} - 2 \dot{\alpha})(2\dot{\beta}_{11} + \dot{\alpha}) + (\dot{\beta}_{22} - 2 \dot{\alpha})(2\dot{\beta}_{22} + \dot{\alpha}) + \nonumber\\
(\dot{\beta}_{33} - 2 \dot{\alpha})(2\dot{\beta}_{33}  
+ \dot{\alpha}) - {1\over 2}[3\dot{\beta}^2_{33} + (\dot{\beta}_{11} - \dot{\beta}_{22})^2 -12\dot{\alpha}^2]\} + \nonumber\\
{N\over 2}e ^{2\alpha}Tr(2e ^{-2\beta} -e ^{4\beta}).
\end{eqnarray}}

The evolution of a quantum model of the universe may be described \'a la Halliwell [10] by the following propagator 

{\setlength\arraycolsep{2pt}\begin{eqnarray}
U(q'', q') = (\tau '' - \tau ')\int d[N^{\bot}]\delta(\dot{N}^{\bot})d[p_l]d[q^l]exp\{i\int d\tau [p_l\dot{q}^l - N^{\bot}H_{\bot}]\},
\end{eqnarray}}

\noindent where $q'' = (\alpha'', \beta'', \tau'')$ and $q' = (\alpha', \beta', \tau')$.

From this last expression we may evaluate the probability transition $P = \vert U \vert^2$ from the initial configuration $q'$ to the final $q''$.

 For our metric we may write the propagator as
  
{\setlength\arraycolsep{2pt}\begin{eqnarray}
U(q'', q') = (\tau '' - \tau ')\int dNd[p_+]d[p_-]d[p_{\alpha}]d[\alpha]d[\beta_+]d[\beta_-]exp\Bigl[i\int_ {\tau '}^{\tau ''}\{p_+\dot{\beta}_+ + p_-\dot{\beta}_- \nonumber\\
- p_{\alpha}\dot{\alpha} - {N\over 3\pi}e ^{-3\alpha}(p^2_+ + p^2_- - p^2_{\alpha}) + {3\pi N\over 2}e ^{\alpha - 2\beta_+}(e ^{-\sqrt{12}\beta_-} + e ^{\sqrt{12}\beta_-} \nonumber\\ 
+ e ^{6\beta_+}) - {3\pi N\over 4}e ^{\alpha + 4\beta_+}(e ^{-4\sqrt{3}\beta_-} + e ^{4\sqrt{3}\beta_-} + e ^{-12\beta_+})\}d\tau\Bigr],
\end{eqnarray}}

\noindent where $\beta_- = {1\over 2\sqrt{3}}(\beta_{11} - \beta_{22})$, $\beta_+ = {1\over 2}(\beta_{11} + \beta_{22})$ and $\beta_{33} = -2\beta_+$ [9].

First the integrals with respect to the momenta will be carried out, and in this integration we use the result $\int d[p]exp\{- {1\over 2}([p], A[p]) + ([q], [p])\} = exp\{{1\over 2}([q], A^{-1}[q]) \}$ [6], where $([q], [p]) = \int_ {\tau '}^{\tau ''}q(\tau)p(\tau)d\tau$.

{\setlength\arraycolsep{2pt}\begin{eqnarray}
\int d[p_+]d[p_-]d[p_{\alpha}]exp\Bigl[i\int_ {\tau '}^{\tau ''} d\tau \{p_+\dot{\beta}_+ + p_-\dot{\beta}_-  - p_{\alpha}\dot{\alpha}  \nonumber\\   
+ {N\over 3\pi}e ^{-3\alpha}(p^2_{\alpha} - p^2_+ - p^2_-)\}\Bigr] =   exp\Bigl[{3i\pi\over 4N}\int_ {\tau '}^{\tau ''}\{\dot{\beta}^2_+ + \dot{\beta}^2_- - \dot{\alpha}^2\}e ^{3\alpha}d\tau\Bigr].
\end{eqnarray}}

Then the propagator becomes 

{\setlength\arraycolsep{2pt}\begin{eqnarray}
U(q'', q') = (\tau '' - \tau ')\int dNd[\alpha]d[\beta_+]d[\beta_-]exp\Bigl[i\int_ {\tau '}^{\tau ''}\{{3\pi N\over 2}e ^{\alpha-2\beta_+}\nonumber\\
\times (e ^{-\sqrt{12}\beta_-} +  e^{\sqrt{12}\beta_-} +  e ^{6\beta_+}) -  {3\pi N\over 4}e ^{\alpha + 4\beta_+}(e ^{-4\sqrt{3}\beta_-} + e ^{4\sqrt{3}\beta_-} + e ^{-12\beta_+}) \nonumber\\  
+ {3\pi\over 4N}e^{3\alpha}(\dot{\beta}^2_+ + \dot{\beta}^2_- - \dot{\alpha}^2)\}\Bigr]d\tau. 
\end{eqnarray}}

In order to obtain an analytical expression for our propagator let us now consider a more symmetric case, we will introduce two restrictions $\beta_- = 0$ and $0 < \vert\beta_+ \vert \ll 1$. 

The first of these two conditions means small but not vanishing anisotropy. Hence, the resulting Hamiltonian acquires a very simple form [11] and in consequence our propagator becomes (from now on we drop the subindex of $\beta_+$)

{\setlength\arraycolsep{2pt}\begin{eqnarray}
U(q'', q') \cong (\tau '' - \tau ')\int dNd[\alpha]d[\beta]exp\Bigl[i\int_ {\tau '}^{\tau ''}\{{3\pi N\over 4}e ^{\alpha}(1 - 8\beta^2) \nonumber\\ 
+ {3\pi\over 4N}e^{3\alpha}(\dot{\beta}^2 - \dot{\alpha}^2)\}d\tau\Bigr].
\end{eqnarray}}

As an additional approximation we will take only terms up to second order in $\alpha$, namely $e^{\alpha} \cong 1 + \alpha + {\alpha^2\over 2}$. 

We now evaluate the restrictions that this approximation imposes on the propagator.

Clearly, the path integrals appearing in expression (8)  may be rewritten as follows

{\setlength\arraycolsep{2pt}\begin{eqnarray}
\int d[\alpha]d[\beta]exp\Bigl[i\int_ {\tau '}^{\tau ''}\{{3\pi N\over 4}e ^{\alpha}(1 - 8\beta^2) 
+ {3\pi\over 4N}e^{3\alpha}(\dot{\beta}^2 - \dot{\alpha}^2)\}d\tau\Bigr]  =  \nonumber\\ 
 \int d[\alpha] exp\Bigl[i\int_ {\tau '}^{\tau ''} ({3\pi N\over 4}e ^{\alpha}  - {3\pi\over 4N}e^{3\alpha}\dot{\alpha}^2) d\tau\Bigr] \times\nonumber\\ 
 \int d[\beta] exp\Bigl[i\int_ {\tau '}^{\tau ''}  (-6\pi Ne ^{\alpha}\beta^2 + {3\pi\over 4N}e^{3\alpha}\dot{\beta}^2)d\tau
\Bigr].
\end{eqnarray}}

Let us now analyze the path integral on $[\beta]$. It is readily seen that

{\setlength\arraycolsep{2pt}\begin{eqnarray}
\int d[\beta] exp\Bigl[i\int_ {\tau '}^{\tau ''}  (-6\pi Ne ^{\alpha}\beta^2 + {3\pi\over 4N}e^{3\alpha}\dot{\beta}^2)d\tau  \Bigr]   = exp [i{3\pi \over 4N}e ^{3\alpha}\beta \dot{\beta}] \mid _ {\tau '}^{\tau ''} \times \nonumber\\ 
\int d[\beta] exp\Bigl[i\int_ {\tau '}^{\tau ''}  [-6\pi Ne ^{\alpha}\beta^2 - {3\pi\over 4N}\beta {d\over d\tau}(e ^{3\alpha}{d\beta \over d\tau})]d\tau  \Bigr].  
\end{eqnarray}}

But, if we define the operator $ A = 12\pi iNe^{\alpha} + {3\pi\over 2N}i{d\over d\tau}(e^{3\alpha}{d\over d\tau}) $, then we may rewrite  the path integral appearing on the right--hand side of (10) as follows

{\setlength\arraycolsep{2pt}\begin{eqnarray}
\int d[\beta] exp\Bigl[i\int_ {\tau '}^{\tau ''}  [-6\pi Ne ^{\alpha}\beta^2 - {3\pi\over 4N}\beta {d\over d\tau}(e ^{3\alpha}{d\beta \over d\tau})]d\tau  \Bigr]  = \nonumber\\
\int d[\beta] exp\Bigl[-{1\over 2}([\beta], A[\beta])  + ([c], [\beta])\Bigr],
\end{eqnarray}}

\noindent where we have $c = 0$ and $([q], [\beta] ) = \int_ {\tau '}^{\tau ''} q(\tau) \beta (\tau) d\tau$, (see [6] pag(45) ).

Therefore, we may evaluate (11).

{\setlength\arraycolsep{2pt}\begin{eqnarray}
\int d[\beta] exp\Bigl[i\int_ {\tau '}^{\tau ''} [-6\pi Ne ^{\alpha}\beta^2   - {3\pi\over 4N}\beta {d\over d\tau}(e ^{3\alpha}{d\beta \over d\tau})]d\tau  \Bigr]  = exp\Bigl[{1\over 2}([c], A^{-1}[c])\Bigr] = 1.
\end{eqnarray}}

Introducing (12) into (10) we obtain 

{\setlength\arraycolsep{2pt}\begin{eqnarray}
\int d[\beta] exp\Bigl[i\int_ {\tau '}^{\tau ''}  (-6\pi Ne ^{\alpha}\beta^2 + {3\pi\over 4N}e^{3\alpha}\dot{\beta}^2)d\tau  \Bigr]   = exp [i{3\pi \over 4N}e ^{3\alpha}\beta \dot{\beta}] \mid _ {\tau '}^{\tau ''}.
\end{eqnarray}}

Let us now consider the path integral on $[\alpha]$ on the right--hand side of expression (9) and define the operator $A$, such that if  $f: \Re \rightarrow \Re$, then $A(f) = - {3iN\pi\over 2}[{f^2\over 2!} + {f^3\over 3!} + {f^4\over 4!} + ...] -  {3i\pi\over 2N}{d\over d\tau}(e ^{3f}{df\over d\tau})$. 

Proceeding in the same way as before we obtain 

{\setlength\arraycolsep{2pt}\begin{eqnarray}
\int d[\alpha] exp\Bigl[i\int_ {\tau '}^{\tau ''} ({3\pi N\over 4}e ^{\alpha}  - {3\pi\over 4N}e^{3\alpha}\dot{\alpha}^2) d\tau\Bigr]  = \nonumber\\
exp\Bigl[i {3N\pi\over 4}({\tau ''} - {\tau '} ) - i{3\pi \over 4N}(e^{3\alpha}\alpha \dot{\alpha}) \mid _ {\tau '}^{\tau ''}\Bigr] \times 
 exp \Bigl[ {1\over 2}([c],A^{-1}[c]) \Bigr],
\end{eqnarray}}

\noindent here $c = {3i\pi\over 4}N$.

To resume, 

{\setlength\arraycolsep{2pt}\begin{eqnarray}
\int d[\alpha]d[\beta]exp\Bigl[i\int_ {\tau '}^{\tau ''}\{{3\pi N\over 4}e ^{\alpha}(1 - 8\beta^2) 
+ {3\pi\over 4N}e^{3\alpha}(\dot{\beta}^2 - \dot{\alpha}^2)\}d\tau\Bigr]  =  \nonumber\\  
exp\{i {3\pi \over 4N} \Bigl[e^{3\alpha''}(\beta''\dot\beta''  - \alpha''\dot\alpha'') + e^{3\alpha'}( \alpha'\dot\alpha'  - \beta'\dot\beta')\Bigr] \}\times \nonumber\\
exp\{i{3\pi N\over 4}(\tau''-\tau')  + {3\pi iN\over 8} \int _ {\tau '}^{\tau ''} A^{-1} ({3\pi iN\over 4})d\tau \}.
\end{eqnarray}}

This last result allows us to write the propagator as follows

{\setlength\arraycolsep{2pt}\begin{eqnarray}
U(q'', q') = (\tau '' - \tau ')\int dNexp\{ i {3\pi N\over 4}(\tau''-\tau')  + \nonumber\\
i{3\pi \over 4N} \Bigl[e^{3\alpha''}(\beta''\dot\beta''  - \alpha''\dot\alpha'') + e^{3\alpha'}( \alpha'\dot\alpha'  - \beta'\dot\beta')\Bigr]   +  {3\pi iN\over 8} \int _ {\tau '}^{\tau ''} A^{-1} ({3\pi iN\over 4})d\tau     \}.
\end{eqnarray}}

In order  to understand the consequences of our approximation let us now consider the following integral $\int dNexp\{i{3\pi \over 4N} \Bigl[e^{3\alpha''}(\beta''\dot\beta''  - \alpha''\dot\alpha'') + e^{3\alpha'}( \alpha'\dot\alpha'  - \beta'\dot\beta')\Bigr]\}$ and define $a = e^{3\alpha''}(\beta''\dot\beta''  - \alpha''\dot\alpha'') + e^{3\alpha'}( \alpha'\dot\alpha'  - \beta'\dot\beta')$. With other words, let us neglect at this point the last integral on the right--hand side of (16). Then, we obtain [12]

{\setlength\arraycolsep{2pt}\begin{eqnarray}
\int dNexp\{i{3\pi \over 4N} \Bigl[e^{3\alpha''}(\beta''\dot\beta''  - \alpha''\dot\alpha'') + e^{3\alpha'}( \alpha'\dot\alpha'  - \beta'\dot\beta')\Bigr]\}  = Nexp[i{3a\pi \over N}]  - \nonumber\\
i{3a\pi \over N} [Ln({1\over N}) + {1\over 1\cdot 1!}({a\over N}) +  {1\over 2\cdot 2!}({a\over N})^2 + {1\over 3\cdot 3!}({a\over N}) ^3 + ...].
\end{eqnarray}}

Expression (17) may be rewritten as

{\setlength\arraycolsep{2pt}\begin{eqnarray}
\int dNexp\{i{3a\pi \over 4N}\}  =  N +  i {3\pi \over 4} (1 + LnN)a + i {3\pi \over 4N} ( {3i\pi \over 8} -   {1\over 1\cdot 1!} )a^2 +  \nonumber\\
i {3\pi \over 4N^2} ({1\over 3!}({3i\pi \over 4})^2 -  {1\over 2\cdot 2!} )a^3 +... .
\end{eqnarray}}

Employing (18) and the definition of $a$ we may finally write down

{\setlength\arraycolsep{2pt}\begin{eqnarray}
\int dNexp\{i{3a\pi \over 4N}\}  = N + f_1\alpha'  +  f_2(\alpha')^2 + f_3(\alpha' )^3 +...+\nonumber\\
j_1\alpha'' + j_2(\alpha'')^2 + j_3(\alpha'')^3 +...+ k_1\alpha'\alpha'' + k_2(\alpha'\alpha'')^2 + k_3(\alpha'\alpha'')^3 +...,
\end{eqnarray}}

\noindent where the coefficients $f_n$, $j_n$ and $k_n$ are functions of $N$, $\dot\beta''$,  $\dot\beta'$, $\beta''$, $\beta'$, $\dot\alpha''$ and $\dot\alpha'$.

The introduction of the approximation $e^{\alpha}  \approx 1 + \alpha + {(\alpha)^2\over 2}$ implies that only powers of $\alpha$ up to second order are relevant. In order to be consistent with it, we must keep in (19) only those terms that contain powers of $\alpha$ not higher that 2. Therefore, if the end points $\alpha'$ and $\alpha''$ satisfy  $e^{\alpha'}  \approx 1 + \alpha' + {(\alpha')^2\over 2}$ and $e^{\alpha''}  \approx 1 + \alpha ''+ {(\alpha'')^2\over 2}$, then we have an evolution propagator that is very close to the exact one. Otherwise, the introduced approximation renders a propagator which matches with the correct one only up to second order.

Let us now recall that Planck length is defined as $l_p = \sqrt{{G\hbar\over c^3}}$. In Planckian units, which we use here, $G$, $\hbar$ and $c$ are equal to $1$, that means, $l_p = 1$. In the case of our metric, the scale factors are $r_{ij} = e^{\alpha}(e^{\beta})_{ij} \sim e^{\alpha}$. From the argument following expression (19) we see that the approximated propagator is very close to the correct one only if the scale factors associated to the involved end points have the same order of magnitude as Planck length. With other words, the employed aproximation imposes a condition which  asserts that the propagator is very close to the correct evolution propagator in that stage of the universe history in which the scale factors  have the same order of magnitude as Planck length .

Considering this approximation we may now rewrite the expression for the pro\-pa\-ga\-tor

{\setlength\arraycolsep{2pt}\begin{eqnarray}
U(q'', q') = (\tau '' - \tau ')\int dNd[\alpha]d[\beta]exp\Bigl[\int_ {\tau '}^{\tau ''}\{{3i\pi N\over 4}(1 + \alpha + {\alpha^2\over 2}) \nonumber\\
- 6\pi iN\beta^2 + {3i\pi \over 4N}(\dot{\beta}^2 - \dot{\alpha}^2)\}d\tau\Bigr].
\end{eqnarray}}

Let us consider first the integral

{\setlength\arraycolsep{2pt}\begin{eqnarray}
\int d[\alpha]exp\Bigl[\int_ {\tau '}^{\tau ''}\{{3i\pi N\over 4}(1 + \alpha + {\alpha^2\over 2}) - {3i\pi \over 4N}\dot{\alpha}^2\}d\tau\Bigr] = exp\{{3i\pi N\over 4}(\tau '' - \tau ')\}\nonumber\\
\times \int d[\alpha] exp\Bigl[i\int_ {\tau '}^{\tau ''}\{{1\over 2}(-{3\pi \over 2N})\dot{\alpha}^2 - {1\over 2}(-{3\pi \over 2N})({N^2\over 2}){\alpha}^2 + {3\pi N\over 4}\alpha \}d\tau\Bigr].
\end{eqnarray}}

The functional integral on the right--hand side of (21) may be understood as the propagator of a driven harmonic oscillator, with mass $m = -{3\pi \over 2N}$, frequency $\omega = {N\over\sqrt{2}}$ and where the external force is $F(\tau) = {3\pi N\over 4}$.

The ``classical action'' of this system is [13]

{\setlength\arraycolsep{2pt}\begin{eqnarray}
S_{\alpha} = {-3\pi\over 4\sqrt{2}sin\Bigl({N\over\sqrt{2}}(\tau '' - \tau ')\Bigr)}\Bigl[(\alpha''^2   
+ \alpha'^2 )cos\Bigl({N\over\sqrt{2}}(\tau '' - \tau ')\Bigr) \nonumber\\
- 2\alpha''\alpha' -4(\alpha'' + \alpha')sin^2\Bigl({N\over\sqrt{8}}(\tau '' - \tau ')\Bigr) - 4sin^2\Bigl({N\over\sqrt{8}}(\tau '' - \tau ')\Bigr) \nonumber\\
+ {N\over\sqrt{2}}(\tau '' - \tau ')\Bigr].
\end{eqnarray}}

 Therefore
  
{\setlength\arraycolsep{2pt}\begin{eqnarray}
\int d[\alpha] exp\Bigl[\int_ {\tau '}^{\tau ''}\{{3i\pi N\over 4}(1 + \alpha + {\alpha^2\over 2}) - {3i\pi \over 4N}\dot{\alpha}^2\}d\tau\Bigr] = \nonumber\\ 
\sqrt{{3i\over 4\sqrt{2}sin\Bigl({N\over\sqrt{2}}(\tau '' - \tau ')\Bigr)}}exp\{{3i\pi N\over 4}(\tau '' - \tau ') + iS_{\alpha}\}.
\end{eqnarray}}

In a similar way we have that

{\setlength\arraycolsep{2pt}\begin{eqnarray} 
\int d[\beta]exp\Bigl[\int_ {\tau '}^{\tau ''}\{- 6\pi iN\beta^2 + {3i\pi \over 4N}\dot{\beta}^2\}d\tau\Bigr] = \sqrt{{3\over \sqrt{2}isin\Bigl(\sqrt{8}N(\tau '' - \tau ')\Bigr)}} exp\{iS_{\beta}\},
\end{eqnarray}}

\noindent where $S_{\beta}$ is the classical action of a free harmonic oscillator with mass $m = {3\pi \over 2N}$ and frequency $\omega = \sqrt{8}N$.

{\setlength\arraycolsep{2pt}\begin{eqnarray}
S_{\beta} = {3\pi\over \sqrt{2}sin\Bigl(\sqrt{8}N(\tau '' - \tau ')\Bigr)}\Bigl[(\beta''^2 + \beta'^2 )cos\Bigl(\sqrt{8}N(\tau '' - \tau ')\Bigr) - 2\beta''\beta'\Bigr].
\end{eqnarray}}
 
From the last integrations we obtain the propagator of a quantum mixmaster universe with small anisotropy, here self--measurement has not been taken into account. 

{\setlength\arraycolsep{2pt}\begin{eqnarray}  
U(q'', q') = \sqrt{{9\over 8}}(\tau '' - \tau ')\int dN {exp\Bigl[ i({3\pi N(\tau '' - \tau ')\over 4} + S_{\alpha} + S_{\beta})\Bigr]\over \sqrt{sin\Bigl({N\over\sqrt{2}}(\tau '' - \tau ')\Bigr)sin\Bigl(\sqrt{8}N(\tau '' - \tau ')\Bigr)}}.
\end{eqnarray}}

Expression (26) is one of the contributions of this work, namely this propagator was up to now not derived.

We now proceed to introduce self--measurement in our universe. This will be done employing RPIF which is a phenomenological approach. This last fact means that we will introduce some parameters that can not be explained in our model but the approach has the advantage that it allows us to consider the influence of the measuring device and at the same time it also enables us to forget the actual scheme of measurement. 

Self--measurement means that some functions $[\kappa]$, $[\nu]$ and $[\gamma]$ are found as estimates of the corresponding functions $[N]$, $[\beta]$ and $[\alpha]$. 

Invariance under reparametrization $d\tau \rightarrow d\zeta = f(\tau)d\tau$ implies that the weight functionals to be introduced in the path integrals must be invariant under this reparametrization. 

This invariance condition is fulfilled if we consider the following weight functionals 

{\setlength\arraycolsep{2pt}\begin{eqnarray}
\omega_{[\kappa]} = exp\{ -\int_ {\tau '}^{\tau ''}{\vert N - \kappa \vert \over \sigma^2}d\tau\},
\end{eqnarray}}

{\setlength\arraycolsep{2pt}\begin{eqnarray}
\omega_{[\nu]} = exp\{ -\int_ {\tau '}^{\tau ''}{N(\nu - \beta)^2 \over \rho^2}d\tau\},
\end{eqnarray}}

{\setlength\arraycolsep{2pt}\begin{eqnarray}
\omega_{[\gamma]} = exp\{ -\int_ {\tau '}^{\tau ''}{N(\gamma - \alpha)^2 \over \Omega^2}d\tau\}.
\end{eqnarray}}

These terms contain implicitly the interaction between environment and collective variables.

Clearly, we do not know if the self--measurement process of the universe renders these functionals.
 But for a qualitative analysis of the consequences of this self--measurement process in the dynamics of the universe we may neglect in a first a\-ppro\-ach the details in the definition of the involved functionals and therefore we may choose the most convenient functionals. These Gaussian weights lead to Gaussian integrals which can be easily performed.
 
A more precise treatment of this issue demands the analysis of the role that higher multipoles of matter play in the definition of the environment associated with the superspace. From this analysis we could also comprehend how the constants $\rho^2 $, $\Omega^2$ or $\sigma^2$, which in this phenomenological approach can not be explained, are defined by the density fluctuations and gravitational waves present in our universe.

Under this choice expression (20) becomes now 

{\setlength\arraycolsep{2pt}\begin{eqnarray}
U_{[\kappa, \nu, \gamma]}(q'', q') = (\tau '' - \tau ')\int dNd[\alpha]d[\beta ]exp\Bigl[\int_ {\tau '}^{\tau ''}\{{3i\pi N\over 4}(1 + \alpha + {\alpha^2\over 2}) \nonumber\\
- 6\pi iN\beta^2 + {3i\pi \over 4N}(\dot{\beta}^2 - \dot{\alpha}^2) - {\vert N - \kappa \vert \over \sigma^2} - {N(\nu - \beta)^2 \over \rho^2} - {N(\gamma - \alpha)^2 \over \Omega^2}\}d\tau \Bigr].
\end{eqnarray}}

Consider now the expression 

{\setlength\arraycolsep{2pt}\begin{eqnarray}
\int d[\beta ]exp\Bigl[\int_ {\tau '}^{\tau ''}\{{3i\pi \over 4N}\dot{\beta}^2 - 6\pi iN\beta^2 - {N(\nu - \beta)^2 \over \rho^2}\}d\tau\Bigr].
\end{eqnarray}}

It may be seen as the propagator of a free harmonic oscillator with mass $m = {3\pi \over 2N}$ and frequency $\omega = \sqrt{8}N$ under continuous measurement of its position $\beta$, such that the function $\nu(\tau)$ is obtained as result of this measurement and the error done in the position measuring is $\Delta\nu = \sqrt{{2\over \vert N(\tau '' - \tau ')\vert}}\rho$.

The propagator of this oscillator is [6, 13]

{\setlength\arraycolsep{2pt}\begin{eqnarray}
\int d[\beta ]exp\Bigl[\int_ {\tau '}^{\tau ''}\{{3i\pi \over 4N}\dot{\beta}^2 - 6\pi iN\beta^2 - {N(\nu - \beta)^2 \over \rho^2}\}d\tau\Bigr] = \nonumber\\
\sqrt{{3\sqrt{1 - {i\over 6\pi\rho^2}}\over \sqrt{2}isin\Bigl(\sqrt{8}N\sqrt{1 - {i\over 6\pi\rho^2}}(\tau '' - \tau ')\Bigr)}}exp \Bigl[-{\vert N(\tau '' - \tau ')\vert\over \rho^2}<\nu^2> +  iS_{\beta}]\Bigr].
\end{eqnarray}}

Here 

{\setlength\arraycolsep{2pt}\begin{eqnarray}
S_{\beta} \cong {3\pi\Gamma\over \sqrt{2}sin\Bigl(\sqrt{8}N\Gamma (\tau '' - \tau')\Bigr)}\Bigl[ (\beta''^2 + \beta'^2 )\nonumber\\
\times cos\Bigl(\sqrt{8}N\Gamma (\tau '' - \tau')\Bigr) - 2 \beta''\beta' - i{\sqrt{8}N(\tau '' - \tau')(\beta''+ \beta') \over 3\pi\rho^2\Gamma}\nu({\tau'' + \tau'\over 2})\nonumber\\
\times sin\Bigl(\sqrt{8}N\Gamma(\tau '' - \tau')\Bigr) + {2N^2(\tau '' - \tau')^2\over 9\pi^2\rho^4\Gamma^2}\nu ({\tau'' + \tau'\over 2})\nonumber\\
\times\nu({\tau'' + 3\tau'\over 4})sin\Bigl(\sqrt{2}N\Gamma(\tau '' - \tau')\Bigr)sin\Bigl({N\over\sqrt{2}}\Gamma(\tau '' - \tau')\Bigr)\Bigr],
\end{eqnarray}}

\noindent $\Gamma = \sqrt{1 - {i\over 6\pi\rho^2}}$, and $<\nu^2> = {1 \over\tau '' - \tau'}\int_ {\tau '}^{\tau ''}\nu(\tau)^2d\tau$ and where we may understand $S_{\beta}$ as the ``classical action'' of a fictitious complex driven oscillator whose mass and frequency are $m = {3\pi \over 2N}$, $\upsilon = \sqrt{8}N\Gamma$, respectively, and where the external force is $F(\tau) = -i{2N\over \rho^2}\nu(\tau)$.

In the case of the integral$\int d[\alpha]exp\Bigl[\int_ {\tau '}^{\tau ''}\{{3i\pi N\over 4}(\alpha + {\alpha^2\over 2})  - {3i\pi \over 4N}\dot{\alpha}^2- {N(\gamma - \alpha)^2 \over \Omega^2}\}d\tau\Bigr] $ the situation resembles the case of expression (31). 

Indeed, we have a harmonic oscillator with mass $m = -{3\pi \over 2N}$, frequency $\omega = {N\over\sqrt{2}}$ and under the influence of the force $F(\tau) = {3\pi N\over 4}$. 
Here the position $\alpha$ is continuously measured, and $\gamma(\tau)$ and $\Delta\gamma = \sqrt{{2\over \vert N(\tau '' - \tau ')\vert}}\Omega$ are the result and involved error in this measurement process, respectively.

The propagator for this harmonic oscillator is also easily calculated 

{\setlength\arraycolsep{2pt}\begin{eqnarray}
\int d[\alpha]exp\Bigl[\int_ {\tau '}^{\tau ''}\{{3i\pi N\over 4}(1 + \alpha + {\alpha^2\over 2}) - {3i\pi \over 4N}\dot{\alpha}^2 -{N(\gamma - \alpha)^2 \over \Omega^2}\}d\tau \Bigr] = \nonumber\\
\sqrt{{3i\sqrt{1 + {i8\over 3\pi\Omega^2}}\over 4\sqrt{2}sin\Bigl({N\over \sqrt{2}}\sqrt{1 + {i8\over 3\pi\Omega^2}}(\tau '' - \tau ')\Bigr)}}exp\Bigl[{3i\pi N\over 4}(\tau '' - \tau ') \nonumber\\
- \vert N(\tau '' - \tau ')\vert{<\gamma^2>\over \Omega^2} + iS_{\alpha}\Bigr].
\end{eqnarray}}

Here we have  

{\setlength\arraycolsep{2pt}\begin{eqnarray}
S_{\alpha} \cong {-3\pi\tilde\omega\over 4\sqrt{2}sin\Bigl({N\over \sqrt{2}}\tilde\omega(\tau '' - \tau ')\Bigr)}\Bigl[(\alpha''^2 + \alpha'^2)cos\Bigl({N\over \sqrt{2}}\tilde\omega(\tau '' - \tau ')\Bigr) \nonumber\\
-2\alpha''\alpha' - 4{(\alpha'' + \alpha')\over\tilde\omega^2}sin^2\Bigl({N\over \sqrt{8}}\tilde\omega(\tau '' - \tau ')\Bigr) \nonumber\\
+ i{8\sqrt{2}N(\tau '' - \tau ')(\alpha'' + \alpha')\over 3\pi\Omega^2\tilde\omega}\gamma({\tau '' + \tau '\over 2})sin\Bigl({N\over \sqrt{8}}\tilde\omega(\tau '' - \tau ')\Bigr) 
\nonumber\\
- 4\tilde\omega^{-4}sin^2\Bigl({N\over \sqrt{2}}\tilde\omega(\tau '' - \tau ')\Bigr) + {N\over \sqrt{2}}{(\tau '' - \tau ')\over \tilde\omega^3}\nonumber\\
+ i{4N^2(\tau '' - \tau ')^2\over 3\pi\Omega^2\tilde\omega^2}[\gamma({\tau '' + \tau '\over 2}) + \gamma({\tau '' + 3\tau '\over 4})]\nonumber\\
\times sin\Bigl({N\over \sqrt{8}}\tilde\omega(\tau '' - \tau ')\Bigr)sin\Bigl({N\over \sqrt{32}}\tilde\omega(\tau '' - \tau ')\Bigr) \nonumber\\
+ {32N^2(\tau '' - \tau ')^2\over 9\pi^2\Omega^4\tilde\omega^2}\gamma({\tau '' + \tau '\over 2})\gamma({\tau '' + 3\tau '\over 4})\nonumber\\ 
\times sin\Bigl({N\over \sqrt{8}}\tilde\omega(\tau '' - \tau ')\Bigr)sin\Bigl({N\over \sqrt{32}}\tilde\omega(\tau '' - \tau ')\Bigr)\Bigr],
\end{eqnarray}}

\noindent where $\tilde\omega = \sqrt{1 + {i8\over 3\pi\Omega^2}}$ and $<\gamma^2> = {1 \over\tau '' - \tau'}\int_ {\tau '}^{\tau ''}\gamma(\tau)^2d\tau$ and here we may understand $S_{\alpha}$ as the ``classical action'' of a fictitious complex driven oscillator whose mass and frequency are $m = -{3\pi \over 2N}$, $\upsilon = {N\over\sqrt{2}}\tilde\omega$, respectively, and where the involved external force is $F(\tau) = {3\pi N\over 4} -i{2N\over \Omega^2}\gamma(\tau)$.

Therefore, the propagator with self--measurement is

{\setlength\arraycolsep{2pt}\begin{eqnarray}  
U_{[\kappa, \nu, \gamma]}(q'', q') = \sqrt{{9\over 8}}\Bigl[(1 + {i8\over 3\pi\Omega^2})(1 - {i\over 6\pi\rho^2})\Bigr]^{1\over 4}(\tau '' - \tau ')\nonumber\\
\times \int{exp\Bigl[iS + N(\tau '' - \tau '){3i\pi \over 4} -\vert N(\tau '' - \tau ')\vert({<\nu^2>\over \rho^2} + {<\gamma^2>\over \Omega^2}) - \int_ {\tau '}^{\tau ''}{\vert N - \kappa \vert \over \sigma^2}d\tau \Bigr]\over \sqrt{sin\Bigl({N\over \sqrt{2}}\sqrt{1 + {i8\over 3\pi\Omega^2}}(\tau '' - \tau ')\Bigr)sin\Bigl(\sqrt{8}N\sqrt{1 - {i\over 6\pi\rho^2}}(\tau '' - \tau ')\Bigr)}}dN,
\end{eqnarray}}

\noindent here $S = S_{\alpha} + S_{\beta}$
\bigskip

\section{Interpretaion of Results.}
\bigskip
\bigskip
In order to obtain in (36) a non--vanishing propagator several conditions must be fulfilled, one of them is that $\kappa(\tau)$ has to be almost a constant. Otherwise the term $\int_ {\tau '}^{\tau ''}{\vert N - \kappa \vert \over \sigma^2}d\tau$ generates an exponential decrease in the integrand of (36). From now on let us consider $\kappa(\tau) = \kappa = const$.

We proceed to define $t = \kappa (\tau'' - \tau')$ and $T = N(\tau'' - \tau')$, then (36) reduces to

{\setlength\arraycolsep{2pt}\begin{eqnarray} 
U_{[t, \nu, \gamma]}(q'', q') = \sqrt{{9\over 8}}\Bigl[(1 + {i8\over 3\pi\Omega^2})(1 - {i\over 6\pi\rho^2})\Bigr]^{1\over 4}\nonumber\\
\times \int{exp\Bigl[iS + T{3i\pi \over 4} -\vert T\vert({<\nu^2>\over \rho^2} + {<\gamma^2>\over \Omega^2}) - {\vert T - t \vert \over \sigma^2} \Bigr]\over \sqrt{sin\Bigl({T\over \sqrt{2}}\sqrt{1 + {i8\over 3\pi\Omega^2}}\Bigr)sin\Bigl(\sqrt{8}T\sqrt{1 - {i\over 6\pi\rho^2}}\Bigr)}}dT.
\end{eqnarray}}

Clearly, $\kappa$ is an estimation of the lapse function $N$. Therefore, a gauge invariant physical time $t$ emerges as consequence of the measurement of the lapse function $N$ by higher mulipoles of matter, and from the form of the metric (1) we see that the physical time $t = (\tau'' - \tau')\kappa$ is indeed an estimation of the duration of the interval $[\tau', \tau'']$, 
while the error done in its measurement has the value $\Delta t = \sigma^2$.

From (37) we see that the propagator will have a non--vanishing value only if several conditions are fulfilled. One of them concerns the distance between $T$ and $t$. With other words, if $T \not \in [t - \sigma^2, t + \sigma^2]$, then the integrand of (37) decays exponentially. 
If (26) is going to be a good approximation for (37), then, among other conditions, we must have $t \ll \sigma^2$ and $\sigma^2 \gg 1$. 

Classically this model suggests that empty space has properties with analogies to an elastic solid and that it resists shear strains [9].

On the other hand, from (34) and (32) we have that (37) may be interpreted as the wave function of a system that consists of an infinite number of subsystems, 
where each one of them has a wave function that is proportional to the multiplication of $exp({3i\pi T\over 4})$ and the product of the wave functions of two damped harmonic oscillators, 
the damping term in each one of the oscillators is $exp[- {\vert T - t \vert \over 2\sigma^2}]$. The frequencies of these two oscillators are $\tilde\omega$ and $\tilde\Gamma$ and their positions suffer the action of a continuous measurement process. 

The integration in (37) indicates that the time at which each one of these subsystems was ``turned on'' is not the same. 
The emerging damping term tells us that if we wish to evaluate at time $t$ the wave function of the whole system, then we need to consider only those subsystems that where ``turned on'' at a time $T$ that differs from $t$ by $2\sigma^2$ or less. 
With other words, those subsystems that were ``turned on'' at a time $T$ such that $\vert T - t \vert > 2\sigma^2$ have at time $t$ an almost vanishing contribution to the wave function of the whole system. 

Looking at expressions (32), (33), (34) and (35) we find the additional conditions that lead us from (37) to (26), namely $\Omega^2$, $\rho^2 \gg 1$, $\rho^2 \gg \sigma^2\vert (\beta '' + \beta ')\vert \tilde{\nu}$, $\rho^4 \gg (\sigma^2\tilde{\nu})^2$, $\Omega^2 \gg \sigma^2\vert (\alpha '' + \alpha ')\vert \tilde{\gamma}$, $\Omega^2 \gg \sigma^4\tilde{\gamma}$, $\Omega^4 \gg (\sigma^2\tilde{\gamma})^2$ and ${<\gamma^2>\over \Omega^2} + {<\nu^2>\over \rho^2}\ll \sigma^{-2}$.

Here we have $\tilde{\nu} = Sup\{\vert\nu(\tau)\vert : \tau \in [\tau', \tau'']\}$, $\tilde{\gamma} = Sup\{\vert\gamma(\tau)\vert : \tau \in [\tau', \tau'']\}$. 

Therefore, the conditions that reduce (37) to (26) become now: $t \ll \Delta t$, $\Delta t \gg 1$, $\Omega^2$, $\rho^2 \gg 1$, $\rho^2 \gg \Delta t\vert (\beta '' + \beta ')\vert \tilde{\nu}$, $\rho^4 \gg (\tilde{\nu}\Delta t)^2$, $\Omega^2 \gg \Delta t\vert (\alpha '' + \alpha ')\vert \tilde{\gamma}$, $\Omega^2 \gg \tilde{\gamma}\Delta t^2$, $\Omega^4 \gg (\tilde{\gamma}\Delta t)^2$, ${<\gamma^2>\over \Omega^2} + {<\nu^2>\over \rho^2}\ll {1\over \Delta t}$ .

If any of these conditions is not fulfilled we can not neglect the measuring process and in consequence instead of Halliwell's propagator we must use (37).

Even the presence of small anisotropy, $\beta_- = 0$, imposes on the analyzed limit geometrical conditions that did not emerge previously. 

In the isotropic universe without cosmological constant [8], the limit of small times and not very accurate self--measurement imposes no restrictions at all on the size of the regions in the 3--Geometry in which Halliwell's propagator is valid.

In the anisotropic case we may find in the limit of small times and not very accurate self--measurement combinations of initial and final points in the minisuperspace that lie outside Halliwell's regime. 
Indeed, this regime demands, among other conditions, the fulfillment of $\vert (\beta '' + \beta ')\vert \ll {\rho^2\over \tilde{\nu}\Delta t}$. 
Clearly isotropy means that $\beta ''$ and $\beta '$ must vanish and therefore the last condition is always in this case satisfied, but in the anisotropic case we might have a measurement process in which ${\rho^2\over \Delta t} \ll \tilde{\nu}$. 

With other words, let us consider $\beta '$ and $\tilde{\nu}$ as fixed and take into account two measurement processes such that their phenomenological parameters satisfy $({\rho_1\over \sigma_1})^2 \ll ({\rho_2\over \sigma_2})^2$. 
Then the size of the neighborhood whose center is $\beta '$ in which Halliwell's propagator is valid is in the measurement process with parameters with subindex 1 much smaller than in the remaining case.
 
This might be reformulated as follows. Even in the case of small times and not very accurate self--measurement anisotropy generates a functional dependence between the size of the neighborhoods in the 3--Geometry in which Halliwell's propagator is valid and the parameters of the measurement process. 
To resume, small times and not very accurate self--measurement does not imply the unrestricted validity, as happens in the isotropic case, of Halliwell's propagator. 

This last result is no surprise at all. Remembering that according to DM $\beta$ plays in this model the role of a collective variable, and that these type of variables interact with the environment rendering decoherence (our phenomenological approach has implicitly already taken into account this interaction in expressions (27, 28, 29, 30).
Then we may expect that with the disappearance of $\beta$ the role played in the dynamics of the universe by the interaction between collective variables and environment will become less important.

Let us now withdraw from the last restrictions the condition $\sigma^2 \gg 1$, but still keeping the other ones. 
With other words, we impose all the mentioned restrictions changing only one, namely  instead of having $\sigma^2 \gg 1$ we now consider the limit $\sigma^2 \ll 1$.

We know that the sequence of functions $\delta_n(x) = {n\over 2}e^{-n\vert x \vert}$ has as limit, when $n \rightarrow \infty$, Dirac's delta [14]. 
Hence, if under the condition $\sigma^2 \ll 1$ the integrand of (37) is almost constant in the neighborhood $\vert T - t \vert \ll \sigma^2$, then we may introduce in the propagator the approximation $exp(- {\vert T - t \vert \over \sigma^2}) \approx 2\sigma^2\delta (T - t)$. 
This substitution enables us to rewrite the propagator of the universe in the presence of self--measurement as  

{\setlength\arraycolsep{2pt}\begin{eqnarray} 
U_t(q'', q') = \sqrt{{9\over 8}}\sigma^2\sqrt{{1\over sin\Bigl({t\over \sqrt{2}}\Bigr)sin\Bigl(\sqrt{8}t\Bigr)}}exp\Bigl\{ {3\pi i\over 4}t \nonumber\\
+ i{3\pi\over \sqrt{2}sin\Bigl(\sqrt{8}t\Bigr)}\Bigl[(\beta''^2 + \beta'^2 )cos\Bigl(\sqrt{8}t\Bigr) - 2\beta''\beta'\Bigr] -\nonumber\\
i {3\pi\over4\sqrt{2}sin\Bigl({t\over\sqrt{2}}\Bigr)}\Bigl[(\alpha''^2   
+ \alpha'^2 )cos\Bigl({t\over\sqrt{2}}\Bigr) 
- 2\alpha''\alpha' -\nonumber\\
4(\alpha'' + \alpha')sin^2\Bigl({t\over\sqrt{8}}\Bigr) - 4sin^2\Bigl({t\over\sqrt{8}}\Bigr) + {t\over\sqrt{2}}\Bigr]\Bigr\}.
\end{eqnarray}}

Clearly, we have a propagator that can be understood as follows: it is proportional to the product of three terms: i) $exp({3\pi i\over 4}t)$, ii) propagator of a free harmonic oscillator with mass $m = {3\pi \over 2}$ and frequency $\upsilon = \sqrt{8}$, iii) the propagator of a driven harmonic oscillator whose mass and frequency are $m = -{3\pi \over 2}$, $\upsilon = {1\over\sqrt{2}}$, respectively, and where the external force is $F(t) = {3\pi \over 4}$.

The scale factors are $r_{ij} = e^{\alpha}(e^{\beta})_{ij}$. The errors in the measurement of $\beta_{ij}$ and $\alpha$ are, approximately, ${\rho\over\sqrt{t}}$ and ${\Omega\over\sqrt{t}}$, respectively. 
Therefore, the errors in the measurement of the scale factors are given by $\Delta r_{ij} \sim r_{ij}{(\rho + \Omega)\over \sqrt{t}}$, 
hence ${\Delta r_{ij}\over r_{ij}} \sim {(\rho + \Omega)\over \sqrt{t}}$, which means that for times smaller than $(\rho + \Omega)^2$ the concept of scale factor is meaningless. 
An increase in the inaccuracy in the measurement of $\alpha$ (or of $\beta$), which is equivalent to an increase of $\Omega$ (or of $\rho$), implies an increase in the size of the time region in which the concept of scale factor is meaningless. 

It is clear that from this phenomenological approach we can not explain $\rho^2 $, $\Omega^2$ or $\sigma^2$. 
Any feasible explanation of them must analyze the role that higher multipoles of matter play in the definition of the corresponding environment. 
\bigskip
\bigskip

\section{Conclusions.}
\bigskip
\bigskip

We have constructed Halliwell's propagator for the case of a Mixmaster universe with small but non--vanishing anisotropy. Afterwards, in the context of the Decoherence Model, we have introduced in this system a self--measurement process, in which higher multipoles of matter act as environment for the superspace variables that in this proposal play the role of collective variables. 

Employing the Restricted Path Integral Formalism we have also calculated Halliwell's modified propagator, which appears as a consequence of this self--measurement process. 
This formalism has enabled us to take into account the influence of the measuring device without knowing the actual scheme of measurement. 

We have shown that a gauge invariant physical time appears as consequence of this self--measurement process. The restrictions that lead us from Halliwell's modified propagator to the usual Halliwell's propagator were also found. 

The validity region of Halliwell's propagator, and in consequence of Wheeler--DeWitt equation, is restricted by the presence of anisotropy. 
There is a set of conditions, not emerging in the isotropic case, that renders a functional dependence between the features of the self--measurement process and the size of the neighborhoods in the 3--Geometry in which Wheeler--DeWitt equation is valid. 

This fact constitutes no surprise at all, the presence of more collective variables means that the interaction Hamiltonian between environment and collective degrees of freedom plays a more decisive role in the dynamics of our universe.

We obtain also an expression for the threshold in time beyond which the scale factors of this model are meaningless, namely for times smaller than $(\rho + \Omega)^2$ we may not speak of scale factors. 

The analysis of the general case (without the use of the approximation $e^{\alpha} \sim 1 + \alpha + {\alpha^2\over 2}$) remains to be done.  But as we here have seen, the emergence of a physical time is a direct consequence of the interaction between environment and collective variables, which in our case are the elements of the spatial metric. We have also seen that an increase in the number of collective variables reduces the validity region of Halliwell's operator, this last statement may be reformulated as follows, if we have more collective variables then the interaction plays a more decisive role in the dynamics of the universe. 

It is readily seen that, if we do not employ the aforementioned approximation, then one of the elements of the spatial metric will not have a finite number of terms (we must in this case use the complete series for $e^{\alpha}$ and not only the first three terms)  and therefore we could expect that the interaction between environment and collective variables in this case will play a more decisive role in the dynamics of the universe (because we will now have more collective variables, the complete series of $e^{\alpha}$) that in the case in which we use the aforementioned restriction and in consequence  a physical time will again emerge.  We could even expect, as a consequence of the presence of more terms in this interaction, that the number of conditions to be satisfied in order to fall into Halliwell's regime will be larger, which implies that the validity region of Halliwell's operator will be smaller than in our case.                                       
\bigskip
\bigskip

\bigskip
 
\Large{\bf Acknowledgments.}\normalsize
\bigskip

A. Camacho would like to thank  A. A. Cuevas--Sosa for his help and D.--E. Liebscher for the fruitful discussions on the subject. 
This work was partially supported by CONACYT Posdoctoral Grant No. 983023. The hospitality of Astrophysikalisches Institut Potsdam is also kindly acknowledged.

\end{document}